\documentclass[twocolumn,usenatbib]{mnras}

\usepackage{graphicx}
\usepackage{amsmath}

\def\pmb#1{\setbox0=\hbox{#1}%
    \kern-.025em\copy0\kern-\wd0
    \kern.05em\copy0\kern-\wd0
    \kern-.025em\raise.0433em\box0}

\def\ltsima{$\; \buildrel < \over \sim \;$}
\def\gtsima{$\; \buildrel > \over \sim \;$}
\def\simlt{\lower.5ex\hbox{\ltsima}}
\def\simgt{\lower.5ex\hbox{\gtsima}}
\def\p2Y{\;_2Y}
\def\m2Y{\;_{-2}Y}

\def\mk2{\mu {\rm K}^2}
\def\Planck{\it Planck\rm}

\def\LCDM{$\Lambda$CDM}

\newcommand{\MB}{M^{SN}_B}
\newcommand{\Mpc}{\text{Mpc}} 
\newcommand{\Hunit}{~\text{km}~\text{s}^{-1} \Mpc^{-1}}

\def\pmb#1{\setbox0=\hbox{#1}%
     \kern-.025em\copy0\kern-\wd0
     \kern.05em\copy0\kern-\wd0
     \kern-.025em\raise.0433em\box0}

\begin{document}

\title[Dark Energy and $H_0$]{Late Time Dynamical Dark Energy and the CMB-Distance Ladder Tension}

\author[George Efstathiou]{George Efstathiou\\
 Kavli Institute for Cosmology Cambridge and 
Institute of Astronomy, Madingley Road, Cambridge, CB3 OHA.}

\maketitle

\begin{abstract} 
 The traditional distance ladder measures the standardized peak absolute magnitude $\MB$ of Type Ia supernovae (SNIa).
 According to the SH0ES\footnotemark collaboration, the distance ladder value of $\MB$ is highly discrepant with the value inferred  from the cosmic microwave background (CMB) assuming the \LCDM\ cosmology. This CMB-distance ladder tension is insensitive to the actual late time expansion history of the Universe. To derive a value of the Hubble constant, $H_0$,  from  $\MB$ requires a cosmological model. However, the magnitude-redshift relation of SNIa provides tight constrains the expansion history. As a consequence, uncertainties in the conversion of  $\MB$ into   $H_0$ have no significant impact on the CMB-distance ladder tension.
 The tentative claim by the Dark Energy Spectroscopic Instrument collaboration  for late time evolution of the dark energy equation of state
does not alter this conclusion. 

\end{abstract}


\footnotetext{SH0ES: SNe, $H_0$, for the Equation of State of dark energy.}
\section{Introduction}
\label{sec:Introduction}

The first results from cosmic microwave background (CMB) anisotropies observed 
by the \Planck\ satellite \citep{Params:2014} revealed a
discrepancy between the value of the Hubble constant, $H_0$, 
determined from fits of the standard six-parameter \LCDM\ model\footnote{Assuming a
spatially flat Universe dominated at the present day by cold dark matter and  a cosmological constant $\Lambda$,  with a power law initial spectrum of adiabatic fluctuations characterized by a scale spectral index $n_s$.}
and the value inferred from the 
Cepheid-based distance ladder measurement  by the
SH0ES
collaboration \citep{Riess:2011}. This discrepancy has become known as the
`Hubble tension' and is discussed in many review articles \citep[see for example] []{Bernal:2016, Shah:2021, Freedman:2021, Tully:2024}.

A recent analysis by the H0 Distance Network Collaboration \cite[][hereafter H0DN]{H0DN:2026}, building on earlier work by  SH0ES, reports a $7.1 \sigma$ discrepancy between their baseline local distance network value (using Cepheids,  tip of the red giant branch (TRGB) and surface brightness fluctuations as distance indicators) of $H_0 = 73.5 \pm 0.81 \Hunit$ and the base \LCDM\ value  determined from the CMB\footnote{H0DN chose the value
$H_0 = 67.24\pm 0.35 \Hunit$ from the combined analysis of Planck+ACT+SPT  reported by \cite{Camphuis:2026}. ACT: Atacama Cosmology Telescope; SPT: South Pole Telescope.}. It should be pointed out that there is
not yet a consensus on the distance ladder measurements. The Chicago-Carnegie Hubble Program (CCHP) consistently reports: (i)  a lower value of $H_0$
in the region of $70 \Hunit$; (ii) a larger error  of $\sim 2 \Hunit$; (iii)
evidence of systematic differences between Cepheid distances and other indicators
such as TRGB and J-region asymptotic giant branch (JAGB); (iv) systematic differences in these methods depending on bandpass \cite[see e.g.][]{Freedman:2023, Uddin:2023,  Freedman:2025, Hoyt:2026}. Each of these points is contested by  H0DN.

This paper is motivated by the claim in \cite{Turner:2026} (hereafter TH26) that constraints on the expansion history from  measurements of baryon acoustic oscillations (BAO) from the Dark Energy Spectroscopic Instrument \citep{DESI:2024, DESI:2025} may affect the interpretation of the Hubble tension.  

SNIa are
central to the classical distance scale measurements. The three rungs of
the distance ladder involve: (i) geometrical distance anchors, with the
three most commonly used being the maser distance to NGC4258 \citep{Reid:2019},
detached eclipsing binaries in the Large Magellanic Cloud \citep{Pietrzynski:2019}, and Gaia parallax measurements of Galactic Cepheids \citep{Riess:2021a, Stiskalek:2026}; (ii) calibration of a standard candle (e.g. normalization of the Cepheid period-luminosity relation, or magnitude of the TRGB) in nearby galaxies that host a SNIa to determine the standardized peak absolute magnitude of SNIa ($M^{SN}_B$); (iii) constraining the expansion history of the Universe with
distant SNIa to infer a value of $H_0$. 

In an earlier paper \citep[][hereafter E21]{Efstathiou:2021}, I stressed that  rungs (i) and (ii) do not involve any
assumptions about cosmology. The distance ladder therefore leads to a value of
$\MB$ that is independent of cosmology.  In the case
of the base \LCDM\ cosmology inferred from the CMB, the expansion rate is so
well constrained that it is possible to determine the standardized absolute magnitude of SNIa to high accuracy using distant supernovae that  participate  in the Hubble flow.
A comparison of this value for $M^{SN}_B$ with the value determined from rung (ii)
of the distance ladder reveals evidence of a `SNIa absolute magnitude tension'.
The statistical significance of this tension  is clearly independent of the actual expansion history of the Universe. 

Rung (iii), however, requires a model for the expansion history and therefore the value of $H_0$ depends on cosmology. Questions then arise concerning:

\noindent
(a) the  accuracy of observational constraints on the expansion history;

\noindent
(b) whether these observations introduce  significant errors in converting measurements of  $\MB$ into a value for $H_0$;

\noindent
(c) whether one should use distance ladder constraints on $\MB$ or $H_0$ to test cosmological models.

\smallskip

E21 pointed out that misleading science conclusions can arise if a SH0ES $H_0$ prior is used to test theoretical models instead of using a prior on $M^{SN}_B$. 
 Many examples of such misleading analyses can be found in the
 literature \citep[see e.g. the review article by][]{diValentino:2021}\footnote{\cite{Benevento:2020, Camarena:2021} and E21. pointed out that 
 large changes in the value of $H_0$ can arise if there is a sudden late-time ($z \simlt 0.02$) change in the dark energy equation of state, which cannot be constrained accurately by SN or BAO data. Such pathological models are not relevant to this paper and will not be discussed further.}. The answer to point (c) is therefore  always to use an observational prior on $\MB$ and to include SN data to constrain the late time expansion history. The rest of this paper is concerned with points (a) and (b).

Section~\ref{sec:SNAM} reviews the determination of $\MB$ from JWST\footnote{James Webb Space Telescope}  measurements
of Cepheids in nearby SN host galaxies. The analysis here is
a simplified version of that presented by H0DN, focussing on the most accurately measured Cepheid distance moduli. Provided that there are no unidentified  systematic errors in the distance ladder measurements, there is a strong  SNIa absolute magnitude tension with the CMB derived \LCDM\ cosmology that is  independent of the actual expansion history of the Universe. Section~\ref{sec:H0} considers the uncertainties in converting $\MB$ into
a value of $H_0$ using the magnitude-redshift relation inferred from  the
 Pantheon+ SN compilation
\citep[][hereafter abbreviated as Pan+]{Scolnic:2022, Brout:2022}.
The Pan+ data tightly constrains  the 
expansion history leading to uncertainties in the value of $H_0$ that
are much smaller than the Hubble tension inferred by SH0ES and H0DN. These results agree with the conclusions of \cite{Dhawan:2020} and E21. 

Section~\ref{sec:DESIH0} investigates the impact of DESI BAO
measurements on the determination of $H_0$. We begin by
reviewing the `inverse distance ladder' 
adding DESI BAO measurements to Pan+ adopting the 
CMB \LCDM\ scale for the sound horizon. As is well known, the inverse distance ladder gives a low value of $H_0$, consistent with the CMB value derived for the \LCDM\ cosmology. Thus,  a resolution of the CMB-distance ladder tension must necessarily involved new physics that we do not yet understand. In the absence of such an understanding, we  assess the impact of evolving dark energy 
from DESI BAO+CMB+Pan+ measurements on $H_0$, assuming that the only difference with the analysis of Sec.~\ref{sec:H0} is
in the model for the late time expansion history of the Universe. 

Section~\ref{sec:Dovekie} discusses how some aspects of the
analysis of Sec.~\ref{sec:H0} change if the Pan+ SN catalogue
is replaced by the Dark Energy Survey-Dovekie compilation \citep[][hereafter DES-Dovekie]{Popovic:2025, Popovic:2026}.

Our conclusions are summarize in Sec.~\ref{sec:conclusions}.

\section{The SNIa Absolute Magnitude Tension}
\label{sec:SNAM}

\begin{table*}

\begin{center}

  \caption{Data used to illustrate the SN absolute magnitude tension. First column lists the host galaxy. Second column gives the anchor calibrated distance modulus $\mu$ determined from JWST measurements of Cepheids as given in \protect\cite{Riess:2023, Riess:2024, Riess:2025}. The fourth column lists the value of the Hubble constant galaxy-by-galaxy, $ H_0^{a_B}$, inferred by
adopting
the intercept of the Pan+ magnitude-redshift relation, $a_B$, of
Eq.~\ref{equ:SH0ES3}.  The fifth column lists the Type Ia SN associated
with each host galaxy as listed in the Pan+ catalogue. The final row
gives the mean values of $\MB$ and $ H_0^{a_B}$. }

\label{tab:data}

\medskip

\begin{tabular}{l|cccl|} \hline 

NAME   &     $\mu$    &   $M_B^{SN}$  &  $H_0^{\rm a_B}$  &  SN \\ \hline
M101   &     $29.13 \pm 0.02$ & $-19.35\pm 0.12$  &   $70.2\pm 3.8$ &2011fe \cr
N5643  &     $30.49  \pm 0.02$ & $-19.26 \pm 0.06$  &   $73.2\pm 2.1$ & 2013aa,  2017cbv \\
N4536  &     $30.92\pm 0.03$ & $-19.37\pm 0.13$  &   $69.6\pm 4.4$ & 1981B \\
N4424  &     $31.05 \pm 0.13$ & $-19.56\pm 0.23$  &   $63.7\pm 6.8$ & 2012cg \\
N1448  &     $31.29  \pm 0.02$ & $-19.20\pm 0.11$  &   $75.2\pm 3.9$ & 2001el, 2021pit \\
N1365  &     $31.31  \pm 0.03$ & $-19.41 \pm 0.10$  &   $68.3\pm 3.1$ & 2012fr \\
N1559  &     $31.37  \pm 0.02$ & $-19.23 \pm 0.09$  &   $74.2\pm 3.1$ &      2005df \\
N2442  &     $31.44  \pm 0.03$ & $-19.21 \pm 0.09$  &   $75.0\pm 3.2$ & 2015F \\
N7250  &     $31.49  \pm 0.05$ & $-19.21 \pm 0.18$  &   $75.0\pm 6.4$ &2013dy \\
N3972  &     $31.70  \pm 0.04$ & $-19.15 \pm 0.10$  &   $76.9\pm 3.7$ & 2011by \\
N4639  &     $31.79  \pm 0.05$ & $-19.34 \pm 0.13$  &   $70.7 \pm 4.4$ & 1990N \\
N5584  &     $31.81  \pm 0.02$ & $-19.01 \pm 0.08$  &   $82.3 \pm 3.2$ & 2007af \\
N2525  &     $31.94  \pm 0.03$ & $-19.21 \pm 0.08$  &   $74.8\pm 2.9$ & 2018gv \\
N5861  &     $32.11  \pm 0.04$ & $-19.17 \pm 0.11$  &   $76.5\pm 4.1$ &  2017erp \\
N3370  &     $32.22  \pm 0.03$ & $-19.28  \pm 0.09$  &   $72.4\pm 3.0$ &  1994ae \\
N3147  &     $32.92  \pm 0.05$ & $-19.06 \pm 0.11$  &   $80.0\pm 4.0$ & 2021hpr,  1997bq,  2008fr-c \\
N5468  &     $32.98  \pm 0.03$ & $-19.03 \pm 0.05$  &   $80.1\pm 2.5$ & 1999cp,  2002cr \\
Mean   &                    & $-19.20 \pm 0.03$                   &   $75.0\pm 1.1$ &         \\ \hline
\medskip
\end{tabular}
\end{center}
\end{table*}

\begin{figure}
	\centering
	\includegraphics[width=85mm, angle=0]{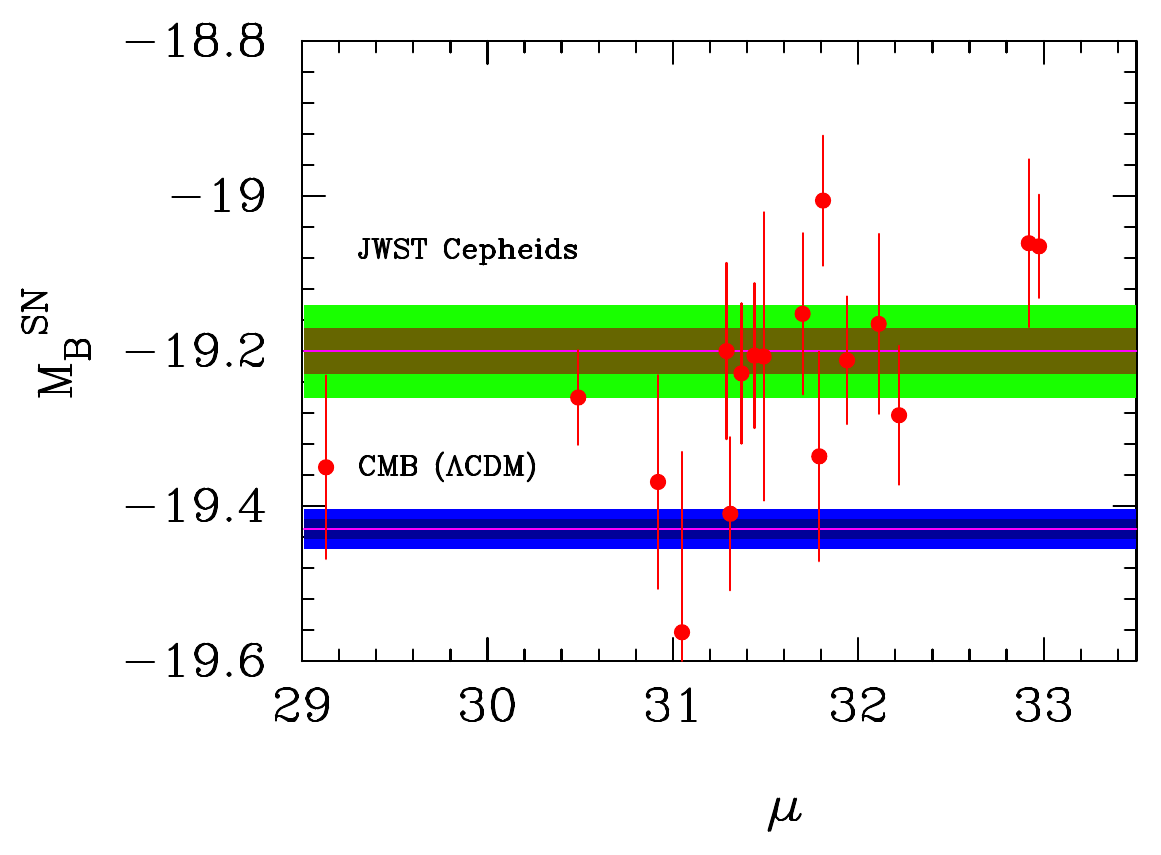} 
	\caption{The points with error bars show the
    peak standardized Type Ia SN magnitude, $\MB$, plotted against  distance modulus, $\mu$ for SN host galaxies  with JWST Cepheid-based distances as listed in Table\ref{tab:data}. In cases where
    galaxies host more than one SN, we plot the mean SN magnitude. The green bands show the $1$ and $2\sigma$ ranges around the mean value of $\MB$ (Eq. \ref{equ:mag3}) allowed by the data points. The blue band show the $1$ and $2\sigma$ ranges of $\MB$ according to \LCDM\ cosmology 
    (Eq. \ref{equ:mag4}) fitted to the Pan+ magnitude-redshift relation over the redshift
    range $0.04 < z^{HD}_{SN} < 1.0$. }

	\label{fig:SNmag}

\end{figure}

In this Section, we present an  analysis of the
SNIa absolute magnitude tension that reproduces the main results of H0DN. We focus on Cepheids in SN host galaxies that have
distance moduli measured with JWST. These measurements are from \cite{Riess:2023, Riess:2024, Riess:2025} and are listed in Table \ref{tab:data}. The last column in this Table lists SNIa
in the host galaxies with photometric data in Pan+.
The third column gives the standardized SNIa absolute magnitude using the Pan+ photometric parameters. (In cases of multiple SN entries in the same host galaxy, the standardized apparent magnitudes 
have been averaged using the covariance matrix provided by Pan+.)  Note that we use the Pan+ catalogue in Secs. \ref{sec:SNAM} and \ref{sec:H0}
because the magnitudes of the nearby SN in Cepheid host galaxies must be analysed using consistent light curve photometry and bias corrections to those of the more distant SN in the  Hubble flow. Various aspects of the Pan+ compilation, principally the model for SN intrinsic scatter and estimates of host galaxy masses,
have changed since it was constructed \citep{Vincenzi:2025}. The recent DES-Dovekie
compilation includes these changes but does not include photometry for the nearby SN with Cepheid distances. Mixing Pan+ photometry of nearby SN with more distant SN from DES-Dovekie can introduce magnitude offsets of $\sim 0.02\ {\rm mag}$ ($\delta H_0 \sim 0.6 \Hunit$). 
The point of view adopted in this paper is that differences in $H_0$ caused by adopting different 
SN compilations should be added to the systematic error budget associated with converting $\MB$ to $H_0$,  so we focus on Pan+. Nevertheless, we attempt to quantify the small differences caused by using DES-Dovekie in place of Pan+ in Sec.~\ref{sec:DESIH0}.

The mean standardized absolute magnitude $\MB$ is determined by minimising
\begin{subequations}.
\begin{equation}
\chi^2_{SN} =  (\MB - m^B_i + \mu_i)C^{-1}_{ij} (\MB - m^B_j + \mu_j), \label{equ:mag1}
\end{equation}
where  $m^B_i$ is the standardized SN peak apparent magnitude of SNIa in galaxy $i$,  and  $C_{ij}$ is an approximation to the covariance matrix of $m^B_i + \mu_i$
given by
\begin{equation}
 C_{ij} =  \left. \begin{array}{ll} (\delta m^B_i)^2 + (\delta \mu_i)^2 +                             (\delta \mu_{\rm anchor})^2&\ i = j, \\
                            (\delta \mu_{\rm anchor})^2  &  \ i\ne j, \end{array} \right\}      \label{equ:mag2}
\end{equation}                            
\end{subequations}
\noindent
where $\delta m^B_i$ and $\delta\mu_i$ are respectively the errors on the standardized SNIa peak apparent magnitudes and JWST distance moduli. The off-diagonal term in Eq.~\ref{equ:mag2} accounts for the correlated error in the first rung geometric anchors, which we take to be the $\delta \mu_{\rm anchor} = 0.02 \ {\rm mag.}$ which is the uncertainty quoted by \cite{Riess:2021} from the combination of using the maser distance modulus to NGC4258, GAIA and HST parallax measurements of Milky Way Cepheids detached eclipsing binaries distances
to the LMC. (The baseline solution in H0DN also uses detached eclipsing binaries in the SMC, but the error in this distance is larger than for the LMC and has an additional error associated with the larger metallicity correction for the Cepheid period-luminosity relation.)

With these approximations, the mean standardized SNIa absolute magnitude averaged over the 
17 galaxies listed in Table~\ref{tab:data} is:
\begin{equation}
\MB = -19.204 \pm 0.030, \qquad {\rm (Cepheids)}.  \label{equ:mag3}
\end{equation}
The individual values of $\MB$ are plotted against distance modulus in Fig.~\ref{fig:SNmag} together with the $1\sigma$ and $2\sigma$ bands from Eq.~\ref{equ:mag3}.
Notice that there appears to be a trend in Fig.~\ref{fig:SNmag} for nearby SN
to have lower values of $\MB$ than more distant ones. This trend has been noted previously, using JWST and HST Cepheid calibrations (see Figure B1 of \cite{Freedman:2021}). \cite{Riess:2024} have argued that sample selection is
partly responsible for the differences between the SH0ES and CCHP calibrations of $\MB$, since SH0ES include more SN at large distances. 

The result of Eq.~\ref{equ:mag3} agrees with the value $\MB = -19.214 \pm 0.037$
of E21 using three distance anchors and the HST Cepheid sample of \cite{Riess:2021}.
The baseline solution of H0DN, which uses Cepheids, TRGB, JAGB, Mira variables,
surface brightness fluctuations, maser galaxies in the Hubble flow gives $\MB = -19.252 \pm 0.022$. These results  illustrate the consistency of the distance ladder calibration (as analysed by H0DN)  between distance indicators and improvements in resolution afforded by JWST compared to HST. 

To compare with the CMB, we use a likelihood similar to the PACT likelihood of \citep{Thibaut:2025}
combining \Planck\ and SPT but using the \Planck\ momento TTTEEE likelihood at low multipoles \citep[described by][]{deBelsunce:2021}, which gives a slightly higher reionization optical depth compared
to the \Planck\ {\tt SimBaL} likelihood \citep{Delouis:2019,Pagano:2020}. Fitting
to the CMB data and the  Pan+ magnitude redshift-relation over the redshift range 
$0.04 < z < 1.0$  we find:
\begin{equation}
\MB = -19.430 \pm 0.013, \qquad (\Lambda{\rm CDM}). \label{equ:mag4}
\end{equation}
The estimates of Eqs.~\ref{equ:mag3} and \ref{equ:mag4} differ by 0.23 mag. at a significance level of $\sim 7 \sigma$ illustrating the SN absolute magnitude tension. No assumptions about the expansion history of the Universe are required to determine
the distance ladder calibration of Eq.~\ref{equ:mag3}. For the estimate of  Eq.~\ref{equ:mag4}, the cosmological parameters of the \LCDM\ model are tightly
constrained by the CMB\footnote{Note that for this CMB data combination we find $H_0 = 68.04 \pm 0.46 \Hunit$
and $\Omega_m = 0.3052 \pm 0.0063$.} leading to the small quoted error. 

As long as the data used in the distance ladder determination of $\MB$  are free of systematic errors\footnote{Including the assumption that the nearby SN in Table~\ref{tab:data} are a representative sample compared to the more distant SN used to compute Eq.~\ref{equ:mag4}.},  then  the  SN absolute magnitude tension must involve new physics beyond that assumed in the base \LCDM\ cosmology.

\section{Inferring $H_0$}
\label{sec:H0}

\begin{table*}
\begin{center}
\caption{The second and third columns give the values of $H_0$,
$q_0$ and $j_0$ fitting the Pan+ magnitude-redshift relation to the
 cosmographic expansion of Eq.\ref{equ:SH0ES2} over two redshift ranges imposing the
 JWST Cepheid value of $\MB$ of Eq.\ref{equ:mag3}.  $H_0$ is given in units of $\Hunit$.The fourth column shows results of an inverse distance ladder fitting the DESI BAO data and Pan+ using the CMB \LCDM\ value of the sound horizon. The derived 
 value of $\MB$ is in very good agreement with Eq.\ref{equ:mag4}. The last column shows fits to the Pan+ magnitude-redshift relation in the $w_0, w_a$ cosmology imposing the DESI+CMB prior of Eqs.~\ref{equ:QCMB1a} and \ref{equ:QCMB1b}. The results in this table use  the {\tt MULTINEST} nested sampler \citep{Feroz:2009, Feroz:2011}.}
\begin{small}

\label{tab:parameters}
\begin{tabular}{c|c|c|c  |c|c| } \hline 
param  &  \multicolumn{2}{c|} {Pan+ \ (Cepheid scale)} & DESI+Pan+  (CMB scale)     &  param &DESI+CMB($w_0w_a$)+Pan+ (Cepheid scale)  \\ \hline
         & $0.023<z_{SN}^{HD}<0.15$  & $0.04 < z_{SN}^{HD}<0.30$  & $0.04 < z_{SN}^{HD}<0.30$   &   &  $0.04 < z_{SN}^{HD}<1.0$ \\
$H_0$    &   $75.0 \pm 1.2$ &  $75.2 \pm 1.2$  &  $68.89 \pm 0.50$&   $H_0$ & $74.7 \pm 1.1$\\
$a_B$     &  $ 0.7162 \pm 0.0025$ &  $0.7172 \pm 0.0030$ &  $0.7182 \pm 0.0007$&  $a_B$ & $0.7145 \pm 0.0025$\\
$q_0 $    &   $-0.53 \pm 0.14$ &  $-0.51\pm 0.09$  & $-0.553 \pm 0.013$ & $\Omega_m$ & $0.309 \pm 0.007$\\
$j_0 $    &   $1.0 \pm 0.6$ &  $1.0 \pm 0.6$ &  $1.24 \pm 0.48$ & $w_0$ &$-0.866\pm 0.068$\\
\ \ $\MB$     &    -            &   -            &  $-19.400 \pm 0.015$ &  $w_a$ & $-0.41 \pm 0.24$ \\ \hline

\end{tabular}
\end{small}
\end{center}
\end{table*}

To convert Eq.~\ref{equ:mag3} into a value of the Hubble constant requires a model for the expansion history of the Universe. The approach adopted by H0DN, and earlier SH0ES
papers, is to fit the SNIa magnitude-redshift relation using a cosmographic
expansion of the luminosity distance with redshift. The standardized SNIa
apparent magnitude is related to absolute magnitude by
\begin{subequations}
\begin{eqnarray}
m_B &= & M_B + 25 + 5 \log_{10} D_L(z),  \label{equ:SH0ES1a}\\
  & = & -5a_B + 5 \log_{10} c \hat d_L(z) \label{equ:SH0ES1b}
\end{eqnarray}
\end{subequations}
where the luminosity distance $D_L$ is in units of Mpc\footnote{We assume a spatially flat geometry throughout this paper and use the `Hubble diagram' redshifts listed in Pan+.}.  In  (\ref{equ:SH0ES1b}), $a_B$ is the intercept of the magnitude-redshift
relation, $5a_B =-(M_B + 25 - 5\log_{10} H_0)$ and $\hat d_L(z)  = H_0D_L(z)/c$. 
To estimate $H_0$, H0DN determine the intercept of the Pantheon SN magnitude-redshift relation by 
fitting the low redshift expansion to the luminosity distance
\begin{equation}
\hat d_l(z) = z\left [ 1 + (1 - q_0) {z \over 2} - {1 \over 6}(1 - q_0 - 3 q_0^2 + j_0)z^2 \right ], \label{equ:SH0ES2}
\end{equation}
over the redshift range $z = 0.023$ to $z =0.15$,  with the deceleration and jerk parameters {\it fixed} to $q_0=-0.55$ and $j_0 = 1$ (close to the CMB values  $q_0 = 3\Omega_m/2-1 \approx -0.54$, $j_0 = 1$ for the base \LCDM\ cosmology). Fitting Eq.~\ref{equ:SH0ES2} to the Pan+ SN, we find
\begin{equation}
a_B = 0.7164 \pm 0.0017,  \label{equ:SH0ES3}
\end{equation}
in very good agreement with the baseline result in H0DN $a_B = 0.716 \pm 0.002$
(see the row labelled V00 in their Table 11).

The fourth column in Table~\ref{tab:data} lists  $H_0$ for each galaxy
using the value of $a_B$ from Eq.~\ref{equ:SH0ES3},  including the errors associated
with $a_B$ and the geometric distance anchors. The average of these values, accounting for correlations is:
\begin{equation}
H_0  = 75.0 \pm 1.1 \ \Hunit,  \label{equ:SH0ES4}
\end{equation}
$1.5  \Hunit$  higher than the baseline value of H0DN. 
 Note that if we use the H0DN `baseline solution' distance moduli 
for the 17 galaxies in Table~\ref{tab:data}, we find $\MB=-19.24 \pm 0.02$, corresponding to $H^{a_B}_0 = 73.5 \pm 1.1 \Hunit$.  This shift is caused by a
small systematic offset between the Cepheid distance moduli and the baseline
distance moduli,  which include other distance indicators.

Instead of using fixed values of $q_0$ and $j_0$, one can
sample over these variables whilst simultaneously solving for
$H_0$. Columns (2) and (3) in Table~\ref{tab:parameters}
show results fiting to SN from  Pan+ over the redshift ranges $0.023 \le z^{HD}_SN < 0.15$ and $0.04 \le z^{HD}_SN < 0.30$. In the latter case, the inferred value of $H_0$ shifts
upwards by $0.2 \Hunit$. These results are consistent with the
H0DN assertion that the value of $H_0$ is  insensitive
to the fits to the Pan+  SN magnitude-redshift relation. Note that the constraints on
$q_0$ and $j_0$ for both redshift ranges are consistent with the expected CMB values 
according to the base \LCDM\ cosmology. 

Although the shifts in $H_0$ in  Table~\ref{tab:parameters} are relatively small, the value in the third column differs by $1.7 \Hunit$ from the baseline value
of $H_0$ quoted by H0DN. This difference is more than twice their quoted error
of $0.8 \Hunit$ and is caused mainly  by  {\it systematic} differences in
the host galaxy distance moduli mentioned above.  This calls into question whether
the H0DN error budget accurately reflects the contributions from systematic errors.

\section{$H_0$ and evolving dark energy}
\label{sec:DESIH0}

DESI BAO measurements combined with CMB observations and Type Ia SN have been interpreted as providing hints for evolving dark energy \citep{DESI:2024, DESI:2025}. In this Section, we investigate
whether these hints have any bearing on the value of the Hubble constant.

We begin first with an application of the inverse distance ladder, summarized in
the fourth column in Table~\ref{tab:parameters}. Here we use the DESI BAO
data calibrated to the \LCDM\ value of the sound horizon, 
$r_d = 147.47\pm 0.26 \ {\rm Mpc}$, given by the PACT likelihood including Planck and ACT CMB lensing. The BAO data are combined with the Pan+ magnitude-redshift relation using the cosmographic model
of Eq.~\ref{equ:SH0ES2}.  The resulting value of $H_0$ is consistent to
within $\sim 1\sigma$ with the \LCDM\ value inferred from the CMB. The value in Table 4 is also consistent with the DES collaboration's application of the inverse distance ladder \citep[giving $H_0 = 67.19^{+0.66}_{-0.64}\Hunit$][]{Camilleri:2025} using the DES 5Y SN catalogue \cite{Vincenzi:2024} combined with DESI BAO measurements. This illustrates the robustness of the inverse distance ladder to changes in the SN  data \citep[see also][]{Popovic:2026b}. In fact, the value of $H_0$ inferred from the
inverse distance ladder has hardly changed  since the earliest applications  \citep{Percival:2010, Heavens:2014, Aubourg:2015} and remains in serious conflict with the H0DN value of $H_0$. 

\begin{figure}
	\centering
	\includegraphics[width=85mm, angle=0]{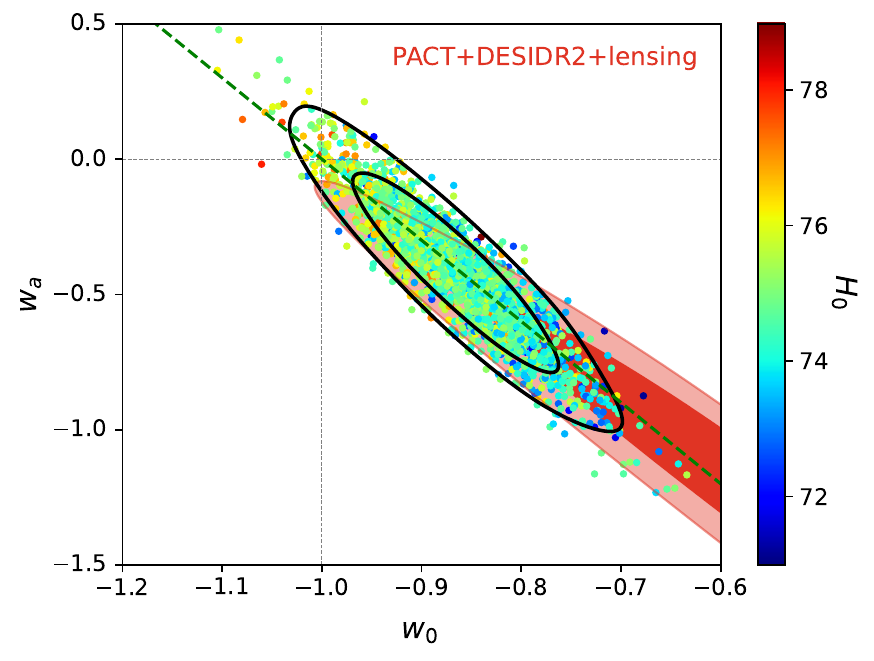} 
	\caption{The red contours show the constraints on $w_0$ and $w_a$ from the CMB combined with DESI DR2 BAO measurements. The CMB constraints use the PACT likelihood combined with CMB lensing (labelled PACT+DESIDR2+lensing).     The solid contours show the $1$ and $2\sigma$ regions allowed
    by the  Pan+ magnitude-redshift relation when the PACT+DESIDR2+lensing constraints on  $\Omega_m$, $w_0$, $w_a$ are imposed as a prior (Eqs.~\ref{equ:QCMB1a} and \ref{equ:QCMB1b}).  The points show samples from the Pan+ chains colour coded by the value of $H_0$ (in units of $\Hunit$).  The dashed green line shows the degeneracy direction $w_a = -3(1-w_0)$ (corresponding to $w(z_{\rm piv}) = -1$  at a pivot redshift of $z_{\rm piv} = 0.5$.}
	\label{fig:DESIH0}

\end{figure}

The red contours in Fig.~\ref{fig:DESIH0} show constraints on the
CPL  parameters $w_0$, $w_a$, where the dark energy equation-of-state 
is assumed to vary with redshift as
\begin{equation}
 w(z) = w_0 + w_a {z \over (1+z)}, \label{equ:CPL}
\end{equation}
\citep{Chevallier:2001, Linder:2003}. These contours use the PACT likelihood+CMB lensing combined with the DESI DR2 BAO measurements
adding  $w_0$ and $w_a$ to the six parameters of the 
base \LCDM\ model. One can see that the 
cosmological constant ($w_0 = -1$, $w_a=0$,  shown by the intersection of the two dotted lines in Fig.~\ref{fig:DESIH0})
lies just outside the $2\sigma$ contours. The degeneracy direction in the $w_0-w_a$ plane tells us that the
CMB+BAO data favour an  angular diameter distance to the
last scattering surface that is consistent with the expectations of the base six parameter \LCDM\ model.
If one then adds the parameters $w_0$ and $w_a$, the preferred models of evolving dark energy belong to the `mirage' family which mimic the base \LCDM\ cosmology \citep{Linder:2007, Lodha:2025}.
In fact, the combination of CMB+BAO agree to high accuracy with the \LCDM\ angular diameter distance at all redshifts
probed by DESI \citep{Efstathiou:2025b}.

TH26 argue that models of  dynamical dark energy favoured
by the DESI collaboration may have a bearing on the Hubble tension. As discussed in Sec. \ref{sec:SNAM}, the SN absolute magnitude tension with \LCDM\ is
independent of the expansion history of the Universe. Nevertheless, one might ask whether evolving dark energy could affect the transformation from SN absolute magnitudes to a value of the Hubble constant. This question, however, is not well posed. The inverse distance ladder results quoted in Table~\ref{tab:parameters} shows that CMB+BAO+SN data favour a 
low value of $H_0$, consistent with the base \LCDM\ CMB value.
Evidently, to achieve consistency with the H0DN distance ladder measurements, a drastic departure of the \LCDM\
model is required (for example, new physics that leads to
a reduction in the sound horizon, see e.g. \cite{Poulin:2023}). However, since  we do not yet have a 
viable model of such new physics,  any assessment of
the impact of evolving dark energy on $H_0$ must necessarily be phenomenological.

One way of proceeding is to impose  priors on the key parameters $\Omega_m$, $w_0$ and $w_a$ that determine the late time expansion history based on the PACT+DESIDR2+lensing
results shown in Fig.\ref{fig:DESIH0}:
\begin{subequations}
\begin{equation}
\pmb{$p$} (\Omega_m, w_0, w_a) = \left ( \begin{array}{c}
\ \ \ 0.3427 \\ -0.517\\ -1.385 \end{array} \right ),  \label{equ:QCMB1a}
\end{equation}
with covariance matrix
\begin{flalign}
{\bf C} =  \left ( \begin{array}{c c c}
    \ 4.0850\times10^{-4} & 3.8608\times10^{-3} & \  -1.0030\times10^{-2}\\
    \ 3.8608\times10^{-3} & 3.9394\times10^{-2} & -0.1066 \\
    -1.0030\times10^{-2} & -0.1066 &\  0.3011 \end{array} \right ).   \label{equ:QCMB1b}   
\end{flalign}
\end{subequations}

Even this approach is not well motivated, however, because 
the DESI BAO data have very little statistical power at redshifts $z<0.15$ used by SH0ES and H0DN to constrain $H_0$.
The exercise therefore amounts to asking : `if the CPL parameterization is 
extrapolated to $z=0$, {\it and} $w_0$ and $w_a$ 
correlate as in Eq.~\ref{equ:QCMB1b}, what is the impact of on $H_0$ if we fit the SN magnitude-redshift relation imposing the distance ladder prior of Eq.\ref{equ:mag3}?'

 The answer to this question is given in the last column of Table~\ref{tab:parameters}. Here we have fitted the Pan+ SN magnitude-redshift relation over the redshift range $0.04< z^{HD}_{SN} < 1.0$ imposing the CMB+DESI
 prior of Eqs.~\ref{equ:QCMB1a} and \ref{equ:QCMB1b}. The resulting constraints on
 $w_0$ and $w_a$ are shown by the solid black contours in  Fig~\ref{fig:DESIH0}.  The points in this plot show samples from chains, colour coded by the value of $H_0$. The mean value is $H_0 = 74.7 \pm 1.1 \Hunit$ differing by only $0.3 \Hunit$ from the $q_0$, $j_0$ determination listed in the second column of Table~\ref{tab:parameters}. The  SN
 data pull the best fit back towards the \LCDM\ model. This behaviour arises because the addition of Pan+ drives $\Omega_m$ back from the
 CMB+DESI BAO value of $\Omega_m = 0.342\pm 0.021$ to
 the base \LCDM\ value of $\Omega_m = 0.300 \pm 0.034$ (for the same data combination). As a consequence, the \LCDM\ model lies within the black $2\sigma$ contour shown in Fig.~\ref{fig:DESIH0}. This illustrates the fragility of the claims for evolving dark energy and their sensitivity to changes in  likelihood combinations and choice
 of SN data.

\section{DES-Dovekie Supernovae}
\label{sec:Dovekie}

The strongest evidence for dynamical dark energy reported by  \citep{DESI:2024, DESI:2025} used the DES 5Y SN catalogue.  The DES-Dovekie catalogue is a revised 
version of DES 5Y incorporating a number of changes including  improved photometric cross-calibration \citep{Popovic:2025, Popovic:2026}. As discussed by \cite{Popovic:2026},
DES-Dovekie  combined with DESI BAO and CMB observations reduces the tension between the $w_0-w_a$ cosmology and \LCDM. Since the inverse distance ladder  is insensitive
to whether one uses the Pan+ or the DES 5Y SN catalogue,  we can anticipate that
using DES-Dovekie in place of Pan+ will not lead to any significant changes to the
conclusions of Sec~\ref{sec:H0}. We test this expectation in this Section.

We first establish a magnitude zero point by requiring $a_B$ to match the Pan+
value of Eq.~\ref{equ:SH0ES3} by fitting DES-Dovekie SN over the redshift range
$0.025-0.15$ to the cosmographic expansion of Eq.~\ref{equ:SH0ES2} with $q_0=-0.55$ and $j_0 = 1$ (i.e. exactly as was done for Pan+). With this calibration, the apparent magnitudes for DES-Dovekie SN are related to the catalogued distances according to
\begin{equation}
m^{\rm DES-Dovekie}_B = \mu^{\rm DES-Dovekie}_B -19.38,
\end{equation}
(in excellent agreement with a recent analysis reported by \cite{Choudhury:2026}).

\begin{figure}
	\centering
	\includegraphics[width=85mm, angle=0]{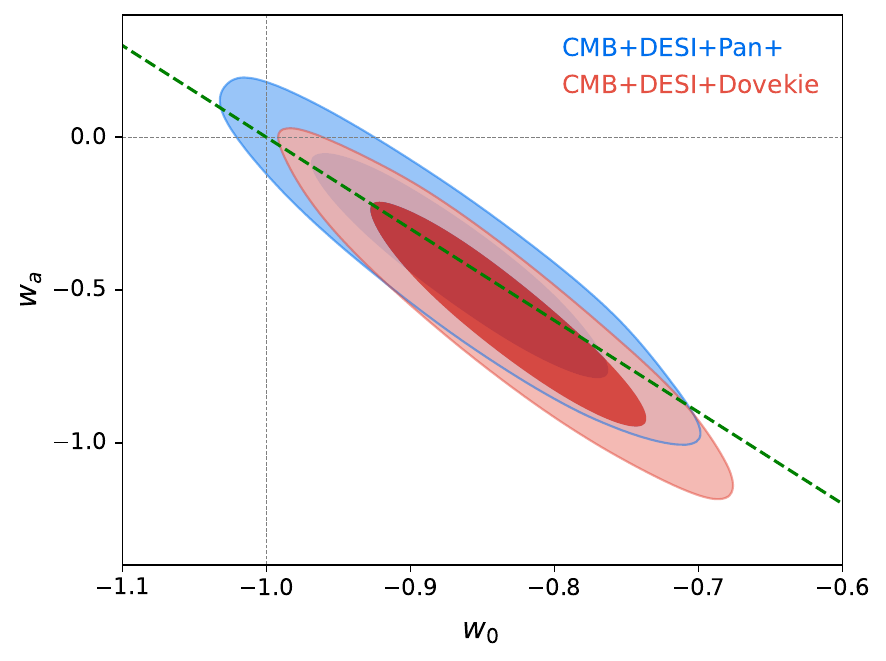} 
	\caption{Constraints on $w_0$ and $w_a$ combining  SN data from Pan+ (blue contours, identical to the solid black contours in Fig.~\ref{fig:DESIH0}) and DES-Dovekie (red contours) with the PACT+DESIDR2+lensing constraints on  $\Omega_m$, $w_0$, $w_a$ (Eqs.~\ref{equ:QCMB1a} and \ref{equ:QCMB1b}).  The dashed green line is as plotted in Fig.~\ref{fig:DESIH0}. }
	\label{fig:DESI_Dovekie}

\end{figure}

We then repeat the analysis of the last column of Table~\ref{tab:parameters}  combining DES-Dovekie SN in the redshift range $0.04< z< 1$ with the  DESI+CMB($w_0w_a$) prior of (Eqs.~\ref{equ:QCMB1a} and \ref{equ:QCMB1b}). The constraints on $w_0, w_a$ are shown in Fig.~\ref{fig:DESI_Dovekie}. DES-Dovekie SN
pull the contours a little further away from \LCDM\ compared to Pan+, but do not strongly exclude a cosmological constant. The DES-Dovekie chains give:
\begin{equation}
    \left. \begin{array}{l}
H_0=  75.1 \pm 1.2 \Hunit,   \\
a_B = 0.7168 \pm 0.0028, \\
\Omega_m  = 0.309 \pm 0.006,  \\
w_0  = -0.833 \pm 0.064,  \\
w_a = -0.58 \pm 0.24. \end{array}
\right\} \label{equ:Dovekie}
\end{equation}
Each of these numbers is consistent with the equivalent Pan+ analysis summarized in
Table~\ref{tab:parameters}. In particular, the Hubble constant changes by only $0.4 \Hunit$.

\section{Conclusions and Discussion}
\label{sec:conclusions}

Distance ladder measurements, as summarized in H0DN, determine the
standardized peak SN absolute magnitude calibrated using local geometric anchors. They do not require a model of the late time expansion history. In the base  \LCDM\ model, the cosmological parameters are so tightly constrained by the CMB that the SN magnitude-redshift relation leads to an accurate determination of the $\MB$
which disagrees by many standard deviations 
from the H0DN value. This absolute magnitude tension is completely independent of the actual expansion history of the Universe and therefore independent of any evidence for evolving dark energy.

SN in the Hubble flow (the third rung of the traditional distance ladder) lead to a determination of $H_0$ that is insensitive to
the redshift range used to constrain the functional form of
the luminosity distance. The uncertainties in $H_0$ associated
with the third rung calibration are an order of magnitude smaller than the
size of the Hubble tension reported by H0DN. 

TH26 claim that the evidence for evolving dark energy
presented by the DESI collaboration has an impact on the Hubble tension. This is not a well posed problem. As the inverse distance ladder demonstrates, reconciling the
H0DN value of $H_0$ with CMB, BAO and SN data requires radical new physics {\it in addition to late time dynamical dark energy}. Such new
physics is very likely to change the interpretation of the CMB.
One cannot therefore  rigorously assess the impact of dynamical
dark energy on the H0DN distance ladder  without a theoretical model that can simultaneously fit the H0DN, CMB, BAO and SN data. To the knowledge of this author,
no such model exists.

Instead, we have adopted a phenomenological approach and constrained the parameters 
$\Omega_m$, $w_0$ and $w_a$ to fit the CMB+DESI BAO data,  treating $w_0$ and $w_a$ as additional parameters to those of the base \LCDM\ cosmology. Adding these constraints to the H0DN measurement of $\MB$ and the Pan+ (or DES-Dovekie)  SN sample, results in a shift to $H_0$ that is small compared to the error quoted by H0DN. Thus, even if one extrapolates the CPL model of
evolving dark energy to low redshifts that are weakly constrained by DESI 
(an assumption which to this author seems extraordinarily poorly motivated)  there is little  
impact on the value of $H_0$.

What then of the `Hubble tension'?  We agree with the authors of H0DN that 
it is extremely unlikely that a single systematic error, localised to one part of the distance latter, can explain the Hubble tension. It is important, however, to 
consider the possibility that several small systematic effects add cumulatively. If no such effects can be found, then  we must be missing new physics. However, any such new physics must be considerably more exotic than the tentative hints of evolving dark energy
 claimed by the DESI collaboration.

\section*{Acknowledgements} 

I thank Roger de Belsunce for allowing me to use 
chains created for a different project.  I am grateful to Suhail Dhawan, Michael Turner and Dragan Huterer for
comments on an earlier draft of this paper. I also than Brodie Popovic for answering questions related to DES-Dovekie.

\section*{Data Availability} 

No new data were generated or analysed in support of this research.

\bibliographystyle{mnras}
\bibliography{H0tension} 
\end{document}